\documentclass{aa}  

\usepackage{soul}
\usepackage{graphicx}
\usepackage{amsmath}
\usepackage{txfonts}
\usepackage{hyperref}
\hypersetup{
    colorlinks,
    linkcolor={blue},
    citecolor={blue},
    urlcolor={blue}
}

\begin{document} 
\title{The challenge of measuring the phase function of debris disks}
\subtitle{Application to HR\,4796}

\author{J. Olofsson\inst{\ref{inst:IFA},\ref{inst:NPF}}
          \and
          J. Milli\inst{\ref{inst:IPAG}}
          \and
          A. Bayo\inst{\ref{inst:IFA},\ref{inst:NPF}}
          \and
          Th. Henning\inst{\ref{inst:MPIA}}
          \and 
          N. Engler\inst{\ref{inst:Zurich}}
          }

\institute{
Instituto de F\'isica y Astronom\'ia, Facultad de Ciencias, Universidad de Valpara\'iso, Av. Gran Breta\~na 1111, Playa Ancha, Valpara\'iso, Chile\\\email{johan.olofsson@uv.cl}\label{inst:IFA}
\and
N\'ucleo Milenio Formaci\'on Planetaria - NPF, Universidad de Valpara\'iso, Av. Gran Breta\~na 1111, Valpara\'iso, Chile\label{inst:NPF}
\and
Univ. Grenoble Alpes, CNRS, IPAG, F-38000 Grenoble, France\label{inst:IPAG}
\and
Max Planck Institut f\"ur Astronomie, K\"onigstuhl 17, 69117 Heidelberg, Germany\label{inst:MPIA}
\and
ETH Zurich, Institute for Particle Physics and Astrophysics, Wolfgang-Pauli-Strasse 27, CH-8093 Zurich, Switzerland\label{inst:Zurich}
}

\abstract{Debris disks are valuable systems to study dust properties. Because they are optically thin at all wavelengths, we have direct access to the absorption and scattering properties of the dust grains. One very promising technique to study them is to measure their phase function, i.e., the scattering efficiency as a function of the scattering angle. Disks that are highly inclined are promising targets as a wider range of scattering angles can be probed.}
{The phase function (polarized or total intensity) is usually either inferred by comparing the observations to synthetic disk models assuming a parametrized phase function, or estimating it from the surface brightness of the disk. We argue here that the latter approach can be biased due to projection effects leading to an increase in column density along the major axis of a non flat disk.}
{We present a novel approach to account for those column density  effects. The method remains model dependent, as one still requires a disk model to estimate the density variations as a function of the scattering angle. This method allows us however to estimate the shape of the phase function without having to invoke any parametrized form.}
{We apply our method to SPHERE/ZIMPOL observations of HR\,4796 and highlight the differences with previous measurements using the surface brightness only, the main differences being at scattering angles smaller than $\sim 100^{\circ}$. Our modelling results suggests that the disk is not vertically flat at optical wavelengths, result supported by comparing the width along the major and minor axis of synthetic images. We discuss some of the caveats of the approach, mostly that our method remains blind to real local increase of the dust density, and that it cannot yet be readily applied to angular differential imaging observations.}
{We show that the vertical thickness of inclined ($\geq 60^{\circ}$) debris disks can affect the determination of their phase functions. Similarly to previous studies on HR\,4796, we still cannot reconcile the full picture using a given scattering theory to explain the shape of the phase function, the blow-out size due to radiation pressure and the shape of the spectral energy distribution, a long lasting problem for debris disks. Nonetheless, we argue that similar effects as the ones highlighted in this study can also bias the determination of the phase function in total intensity.}

\maketitle
%

\section{Introduction}

Dust grains are the building blocks of planets, but there are relatively few ways to accurately characterize their properties. Studies of solar system bodies provide the strongest constraints on the constituents of comets or asteroids (e.g., \citealp{Frattin2019}, \citealp{Bertini2019}), but do not inform us directly on what is taking place during the planet formation stage. Observations of disks around young ($\leq 100$\,Myr old) stars are sensitive to grain sizes that are smaller than typically $100$\,$\mu$m or a few mm. Debris disks, disks of second generation dust, are ideal targets to study dust grains. As those disks are optically thin at all wavelengths, we have direct access to the absorption and scattering properties of the grains, without having to account for non trivial optical depth effects or multiple scattering events. There are two possible ways to characterize dust properties in debris disks, first, via their thermal emission by measuring the spectral slope at (sub-) mm wavelengths (\citealp{Draine2006}, \citealp{MacGregor2016}, constraining the slope of the grain size distribution or the maximum grain size), or modelling their emission features in the mid-IR (\citealp{Olofsson2012}, informing about the dust composition ). The second avenue is to study how stellar light is scattered off of the dust grains (in total intensity or polarized light, at optical or near-infrared wavelengths), either by measuring the colour of the disk between different bands (\citealp{Debes2008}, \citealp{Rodigas2105}), or studying the phase function (e.g., \citealp{Olofsson2016}; \citealp{Milli2017,Milli2019}, \citealp{Ren2019}). Both approaches can bring constraints on the typical grain sizes as well as their porosity. The phase function informs us how efficiently the light is scattered as a function of the scattering angle (between the star, the dust grain and the observer). This approach requires the disk to be spatially resolved and therefore became more popular in the past years with the availability of high angular resolution instruments such as VLT/SPHERE \citep{Beuzit2019} or GPI \citep{Perrin2015}, but pioneering works were led with \textit{Hubble Space Telescope} observations as well (e.g., \citealp{Graham2007}, \citealp{Stark2014}).

\begin{figure*}
\centering
\includegraphics[width=\hsize]{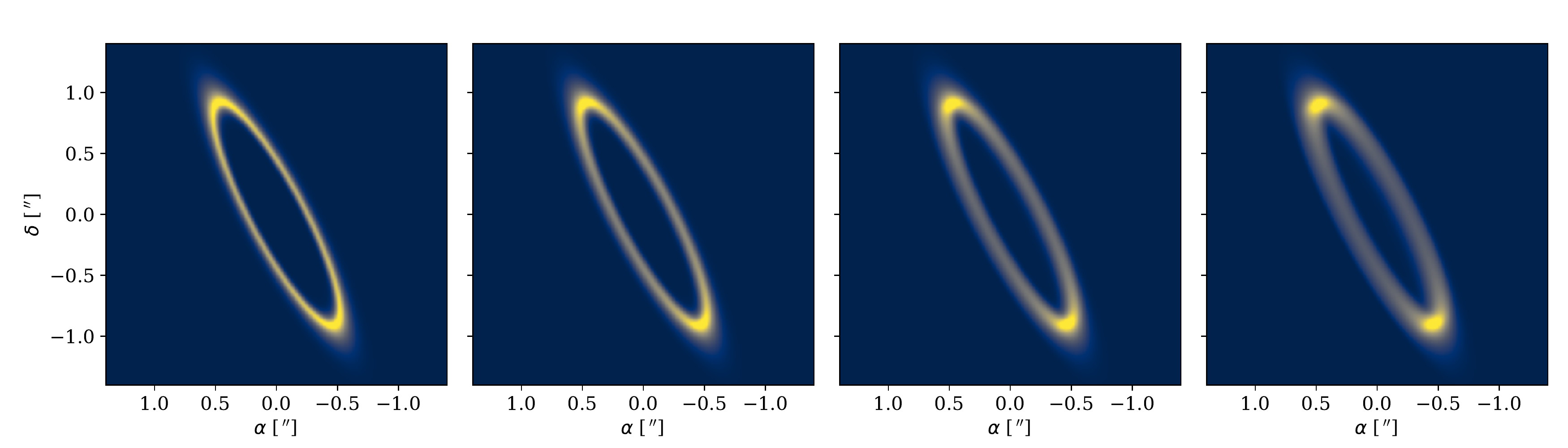}
    \caption{Convolved synthetic images of a debris disk with increasing opening angle. From left to right, $\psi = \mathrm{arctan}(h/r) = 0.01$, $0.03$, $0.05$, $0.07$ (see text for further details).}
\label{fig:opang}
\end{figure*}

There are two different methods to estimate the phase function of debris disks, either from total intensity or polarized light observations, and in this study, we will focus mostly on polarimetric observations. Out of the two approaches, the first one consists of fitting a model to the observations and the phase function is a free parameter of the modelling, either using scattering theory such as Mie or more complex ones (and in that case, the free parameter(s) mostly govern the grain size distribution or the porosity), or using parametrized approximations such as the Henyey-Greenstein one (\citealp{Henyey1941}). The second approach being to measure the surface brightness of the disk as a function of the scattering angle, without the use of a model. The first approach requires a disk model for the dust density distribution as well as a scattering theory (or an approximation) that is able to reproduce the true phase function, while the second approach does not account for changes in column density at different azimuthal angles (as is the case along the semi-major axis of inclined disks). One should note that in this method, the final phase function is not exactly equal to the surface brightness as several correction factors have to be applied (e.g., illumination effects, point spread function dilution, and aperture shape, \citealp{Milli2019}). 

The phase function is best measured for disks with a significant inclination (but not perfectly edge-on as most of the azimuthal information is lost), as a wide range of scattering angles can be probed. Inclinations around $75^{\circ}$ are ideal, but as a consequence, there is an increase in column density along the major axis, if the disk is not infinitely flat. This increase in column density is illustrated in Figure\,\ref{fig:opang} showing disk models computed with different opening angles (increasing from left to right, see Section\,\ref{sec:model} for how the images are computed). The models shown in Fig.\,\ref{fig:opang} all have an isotropic phase function and therefore directly trace the dust column density. One can note that as the opening angle increases, the major axis of the disk becomes brighter compared to the minor axis. Therefore when retrieving the phase function by measuring the surface brightness as a function of the scattering angle, one has to account for the column density variations along the disk. We here present a novel approach to retrieve the phase function in debris disks, in a non-parametric way, but that remains model-dependent on the density distribution throughout the disk.

\section{Determining the phase function}

In this Section, we describe our new approach to determine the phase function in a non-parametric way, and briefly present the observations used to test it beforehand.

\subsection{Observations}

Because of the brightness of the disk around HR\,4796, we use the SPHERE/ZIMPOL polarimetric observations presented in \citet{Milli2019} and \citet{Olofsson2019} to illustrate our method. HR\,4796 is a young ($8\pm2$\,Myr old, \citealp{Stauffer1995}) nearby ($71.9\pm0.7$\,pc, \citealp{Gaia2018}) A-type star surrounded by one of the brightest debris disk ($L_\mathrm{disk}/L_\star \sim 5 \times 10^{-3}$, \citealp{Moor2006}). We used the $Q_\phi$ and  $U_\phi$ images presented in \citet{Milli2019} as the signal to noise ratio is larger close to the semi-minor axis, at the expense of some small artefacts along the semi-major axis. The observations were obtained without a coronagraph and the reader is referred to \citet{Milli2019} for more information on the observing sequence and data processing. For all the calculations described below, we mask all the points that are within $0\farcs35$ from the estimated position of the star and all the pixels that are outside of an elliptical mask with a semi-major axis of $1\farcs25$ (see left panel of Fig.\,\ref{fig:data}) are excluded when computing the goodness of fit.

\subsection{Synthetic disk images}\label{sec:model}

We use an updated version of the code presented in \citet{Olofsson2016,Olofsson2018} which can quickly produce images of (eccentric) debris disks which are not infinitely flat. To summarize briefly, the density distribution is computed as
\begin{equation}\label{eqn:nr}
n(r, z) \propto \left[\left(\frac{r}{r_0}\right)^{-2\alpha_{\mathrm{in}}} + \left(\frac{r}{r_0}\right)^{-2\alpha_{\mathrm{out}}}\right]^{-1/2} \times \mathrm{e}^{-z^2/2h^2},
\end{equation}
where $n$ is the volumetric density, $r_0$ is a reference radius, $\alpha_\mathrm{in}$ and $\alpha_\mathrm{out}$ are the slopes of the density distribution and $h$ is the vertical height of the disk (parametrized with the opening angle $\psi$ such as $\mathrm{tan}\,\psi = h/r$). For eccentric disks, parametrized with two free parameters, the eccentricity $e$ and the argument of pericenter $\omega$, the reference radius $r_0$ depends on the azimuthal angle  $\gamma$ such as
\begin{equation}\label{eqn:r0}
        r_0 = \frac{a(1-e^2)}{1 + e \mathrm{cos}(\omega + \gamma)},
\end{equation} 
where $a$ is the semi-major axis, $\gamma = \mathrm{arctan2}(y,x)$ is the azimuthal angle in the midplane\footnote{We use arctan2 to place the resulting angle in the proper quadrant when both $y$ and $x$ are negative for instance.}, and $x$ and  $y$ are the pixels coordinates (with the origin at the centre of the image) after projecting for the inclination $i$ and rotating according to the position angle $\phi$. To produce images, the code first defines a bounding box, sufficiently large for the volume density to be negligible at the borders. For each pixel of the image, the entry and exit points of the bounding box are calculated, and that ``column'' is divided into $m = 100$ equal parts of the same volume $V$. For each cell, the volume density is computed following Eq.\,\ref{eqn:nr}, the scattering angle $\theta$ is computed from the dot product between the unit vector along the line of sight and the $3$D coordinates at the centre of the cell, and the flux will be the product of the volume density, $V$, and the scattering $S_{11}(\theta)$ (or polarized $S_{12}(\theta)$) phase function. For each pixel, the final flux is the sum over the $m$ cells. The user can provide a $2$D array for $S_{12}$ (or $S_{11}$), a $1$D array for $\theta$ (between $0$ and  $\pi$) and the code will interpolate at the proper scattering angle when computing an image. The array for the phase function is $2$D so that there can be one phase function for the north side and a different one for the south side. Both sides are identified based on the sign of the azimuthal angle (which ranges between $-\pi$ and $\pi$). One should note that the code is not flux-calibrated, therefore, the total dust mass or the polarization degree are not mandatory input parameters. 

To compare the synthetic images to the observations, we follow the approach explained in the Appendix of \citet{Engler2018}. To summarize briefly, the modelled $Q_\phi$ image is decomposed in $Q$ and $U$ images, according to the polar coordinates on the detector. Then both images are convolved with a 2D normal distribution with a full width at half maximum (FWHM) of $34$\,mas ($\sigma = 2$\,pixels, comparable to the observing conditions as reported in \citealp{Milli2019}). The convolved $Q$ and $U$ images are then combined to obtain the final $Q_\phi$ image (see \citealp{Engler2018} for further details). 

\subsection{Description of the approach}

\begin{figure*}
\centering
\includegraphics[width=\hsize]{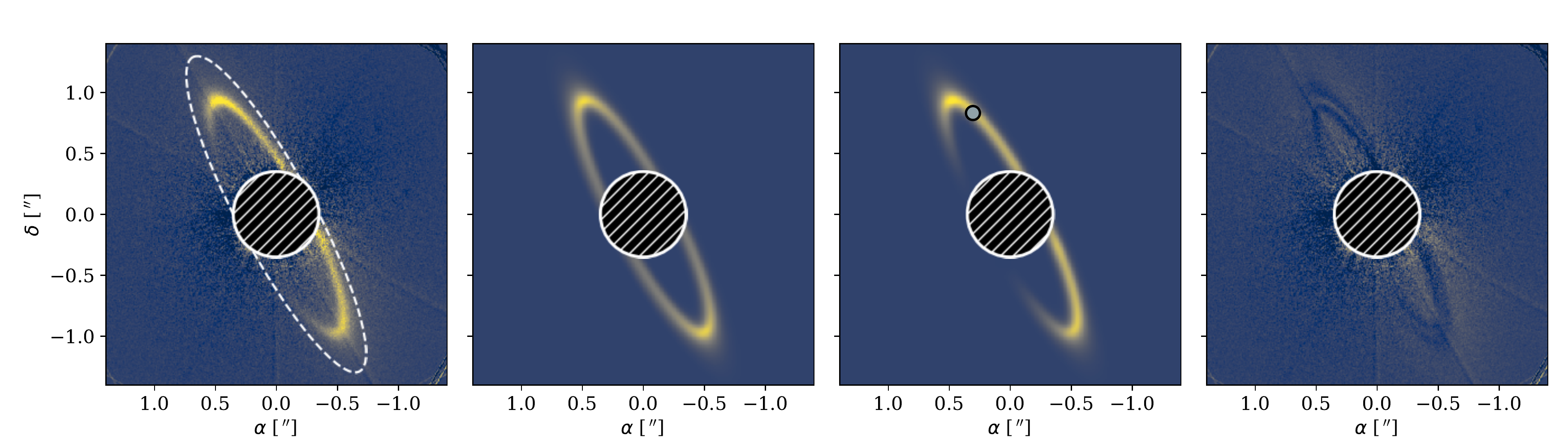}
    \caption{Observations, best-fit model with an istropic phase function, final best-fit model with the revised phase function, and residuals to the SPHERE/ZIMPOL observations of HR\,4796\,A, with the same linear stretch (from left to right). The goodness of fit is estimated between the inner circle and the elliptical mask shown in the leftmost panel. In the central right panel, the location of the pericenter is marked with a circle.}
\label{fig:data}
\end{figure*}

\begin{figure*}
\centering
\includegraphics[width=\hsize]{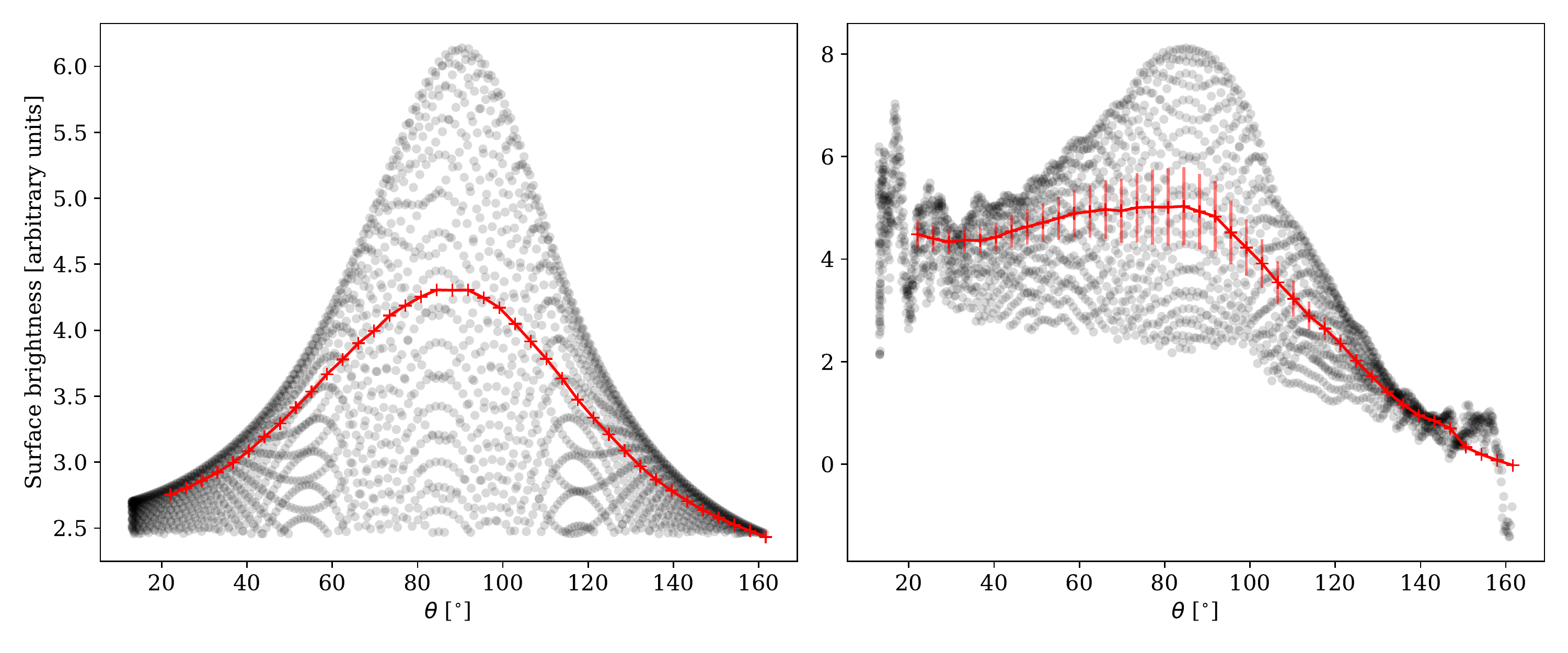}
    \caption{Surface brightness of the north side of the disk, as a function of the scattering angle for the model with an isotropic phase function and for the observations (left and right, respectively). Each black circle represents a pixel. The red curves correspond to the median in each bin of $\theta$.}
\label{fig:surface}
\end{figure*}

The motivation of our approach is to decouple the geometric parameters of the disk (e.g., radius, inclination, eccentricity) from the dust properties (the phase function in this case). This is achieved by comparing a first model computed using an isotropic phase function, which solely traces the dust density distribution, to the observations. By computing the ratio of surface brightness between this first model and the observations one can infer a revised phase function that should account for most of the differences. This revised phase function can then be injected in a new model to be compared to the observations. 

We follow a three-step approach; for a given set of parameters we compute a first model with an isotropic phase function for both north and south sides (with the same number of pixels, and the same pixel scale as the observations, $7.2$\,mas per pixel) to trace the density distribution as a function of the scattering angle. Because the code is not flux-calibrated, we first find the best scaling factor $S_\mathrm{scale}$ that minimizes the difference between the model and the observations, as 
\begin{equation}\label{eqn:scale}
    S_\mathrm{Scale} = \frac{\sum\frac{I_\mathrm{obs} \times I_\mathrm{model}}{\sigma^2}}{\sum\left(\frac{I_\mathrm{model}}{\sigma}\right)^2},
\end{equation}
where $I_\mathrm{model}$ and $I_\mathrm{obs}$ are the images of the model and the $Q_\phi$ image, while $\sigma$ are the uncertainties estimated from the $U_\phi$ image (see next Section, and see Section\,\ref{sec:adi} for a discussion on total intensity observations). The purpose of this first scaling is to try to account for most of the (unknown) multiplicative factors that govern the total flux of the disk (e.g., total dust mass, albedo).

Then, for each side of the disk (north and south) , we estimate the surface brightness of both the model and the observations as a function of the scattering angle in the midplane of the disk. Given that the resulting distribution can be quite noisy for the observations we apply a numerical mask, selecting only the pixels where the model is brighter than $0.45$ times its peak brightness. Given that we use an isotropic phase function for this first model, the whole ring is recovered, for a wide range of scattering angles. 

The second step of the approach is to bin the surface brightness as a function of the scattering angle, for both the model and the observations. The binning is performed over $50$ linearly spaced bins, and for each bin we compute the median value. For the observations, we also compute the median absolute deviation $\sigma_i$ in each bin. Afterwards, we average the resulting distributions using a running mean over $5$ neighbouring bins, to smooth them. For the observed profile, the corresponding uncertainties are estimated as
\begin{equation}\label{eqn:uncer}
    \sigma_\mathrm{binned} = \left( \sum_{i=0}^N \frac{1}{\sigma_i^2} \right)^{-1/2},
\end{equation}
where $N = 5$ is the number of neighbouring bins.

This process is illustrated in Figure\,\ref{fig:surface} (the left panel shows an asymmetry as the disk model is slightly eccentric). Given the mask that excludes the central $0\farcs35$, we are able to probe scattering angles in the range $[22^{\circ}, 163^{\circ}]$ on both the north and south sides. The polarized phase function is estimated from the ratio between the functions representing the observations and the model (propagating the uncertainties at the same time), in other words, the red curve of the right panel divided by the red curve in the left panel of Figure\,\ref{fig:surface}. The scaling factor $S_\mathrm{scale}$ of the isotropic model accounts for most of the differences with the observations but this model is not necessarily a good match, and as a consequence the $y$-axis of Figure\,\ref{fig:surface} may be different. When computing the ratio between the two averaged surface brightness there may still be some contribution from unconstrained ``flux-calibration'' factors and those factors will be incorporated in the resulting phase function, which is therefore a scaled (up or down) version of the true phase function\footnote{One could normalize the phase function over $4\pi$ steradians but since we do not probe the full range of scattering angles, this would remain an approximation}. 

We then compute the final model with the newly evaluated phase function. Once the second model is computed we need to find the new best scaling factor $S_\mathrm{Scale}$ that minimizes the $\chi^2$ following Eq.\,\ref{eqn:scale}. One should note that computing the second model is only necessary to compare the model to the observations and estimate a goodness of fit to find the best parameters of the disk model. 

From Figure\,\ref{fig:surface}, one can see that by estimating the phase function from the observed surface brightness, for a non flat disk, we over-estimate it quite significantly at scattering angles near $90^{\circ}$ (or under-estimate it a smaller angles), especially for highly inclined disks (e.g., \citealp{Olofsson2016}, \citealp{Milli2019}). 

\section{Modelling and results}

\begin{table}
\caption{Free parameters and best-fit results.}
\label{tab:grid}
\centering
\begin{tabular}{lcc}
\hline\hline
Parameters & Prior & Best-fit \\
\hline
$a$ [$\arcsec$]       & [$0.90$, $1.15$]  & $1.066\pm0.001$ \\
$i$ [$^{\circ}$]      & [$70$, $80$]      & $77.60\pm0.06$ \\
$\alpha_\mathrm{out}$ & [$-15$, $-5$]     & $-11.78\pm0.20$ \\
$e$                   & [$0.0$, $0.1$]    & $0.026\pm0.002$ \\
$\omega$ [$^{\circ}$] & [$-180$, $-90$]   & $-147.8\pm5.2$ \\
$\psi$   [$^{\circ}$] & [$0.005$, $0.06$] & $0.035\pm0.001$ \\
\hline
\end{tabular}
\end{table}

To determine most accurately the shape of the polarized phase function, we modelled the observations, trying to constrain the most relevant parameters of the disk. We put a strong emphasis on free parameters that can have an effect on the local increase in column density along the major axis. Therefore, the free parameters are the semi-major axis $a$, the inclination $i$, the outer slope of the density distribution $\alpha_\mathrm{out}$ ($\alpha_\mathrm{in}$ being fixed to $+25$), the eccentricity $e$, the argument of periapsis $\omega$, and the opening angle $\psi$. The position angle of the disk has already been well constrained for this dataset and we therefore use a value of $-152.1^{\circ}$ following the results presented in \citet{Olofsson2019}. The uncertainties are estimated from the $U_\phi$ image, computing the standard deviation in concentric annuli of $2$\,pixel width. Neither the $Q_\phi$ nor the $U_\phi$ images are convolved. Overall, since our model is convolved with a point spread function representative of the observations (following the approach outlined in \citealp{Engler2018}), that illumination effects are naturally accounted for in the model, and that we do not use an aperture to measure the surface brightness profiles, we do not need to apply correction factors such as the ones mentioned in the introduction and described in \citet{Milli2019}. 

\begin{figure}
\centering
\includegraphics[width=\hsize]{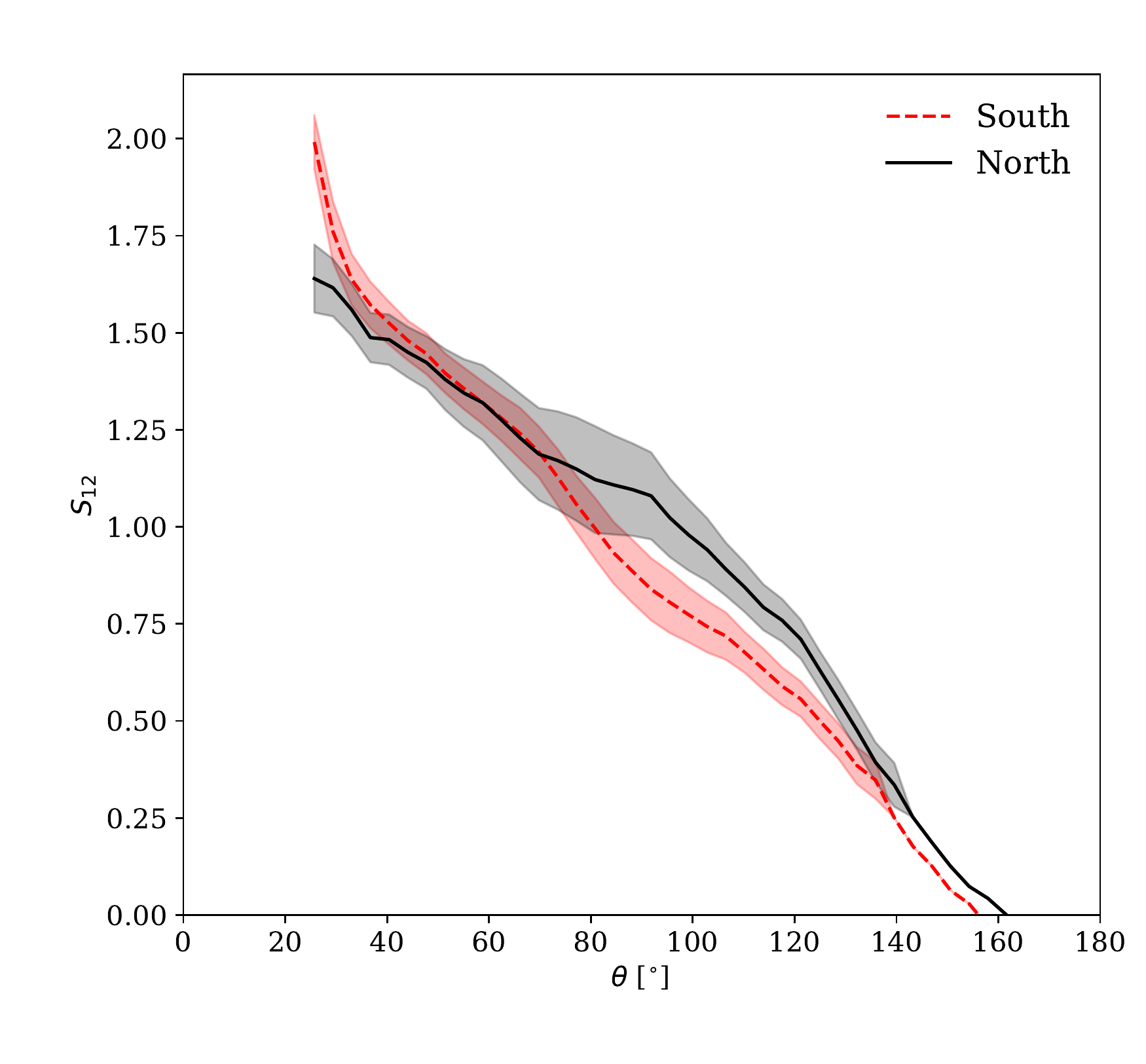}
    \caption{Polarized phase function for the north and south sides (black and dashed red, respectively) as a function of the scattering angle.}
\label{fig:pfunc}
\end{figure}

To identify the most probable solution, we use the \texttt{MultiNest} algorithm (\citealp{Feroz2009}) interfaced to \texttt{Python} via the \texttt{PyMultiNest} package (\citealp{Buchner2014}). The probability distributions are plotted using the \texttt{corner} package \citep{corner} and are presented in Figure\,\ref{fig:corner}. The best-fit values and their uncertainties are reported in Table\,\ref{tab:grid}, while the best-fit model and the residuals are presented in the centre right and right panels of Figure\,\ref{fig:data} (the model with the isotropic phase function is shown in the centre left panel). One should note that the uncertainties reported in Table\,\ref{tab:grid} are most likely under-estimated and should be taken with some caution. This is most likely due to the uncertainties derived from the $U_\phi$ image used to compute the goodness of fit. By measuring the standard deviation in concentric annulii, a strategy commonly used in direct imaging studies, we may be under-estimating the true uncertainties, yielding larger $\chi^2$ values. As a consequence, the Monte-Carlo algorithm may explore a narrower range of values, leading to narrow probability distributions. 
Finally, the polarized phase functions for the north and south sides of the disk are shown in Figure\,\ref{fig:pfunc}, in black and dashed red, respectively.

We find that the semi-major axis is well constrained at $a=1\farcs066 \pm 0.001$, the inclination is $i = 77.60^{\circ}\pm0.06$, the outer slope of the density distribution is $\alpha_\mathrm{out} = -11.8\pm0.2$, the eccentricity $e = 0.026\pm0.002$, the argument of periapsis is  $\omega = -147.8^{\circ}\pm5.2$ (and its location is shown in the middle panel of Fig.\,\ref{fig:data}), and finally, we find that the opening angle is $\psi = 0.035\pm0.001$. From Figure\,\ref{fig:corner}, the degeneracy between $e$ and $\omega$ is clearly noticeable, which explains why we find a smaller value for $e$ compared to the results of \citet{Milli2017,Milli2019}, \citet{Olofsson2019} who found the argument of periapsis to be closer to the projected semi-minor axis of the disk.

\section{Discussion}

\subsection{Residuals and caveats}

The residuals image overall shows that most of the signal from the disk has been removed, especially in the south side of the disk. There is still some signal left in the northern side, and this implies that the smoothed surface brightness profile that we estimated from the observations may not accurately capture the true surface brightness distribution of the northern side. As pointed out in \citet{Olofsson2019}, the brightness asymmetry along the two sides cannot be solely explained by pericenter glow (\citealp{Wyatt1999}) and that there may be an over-density of small dust grains at the north side of the disk. Our modelling approach is blind to local increase in dust density at different azimuthal angles, and there is no easy way around this issue. Nonetheless, the fact that the phase functions are quite similar between the north and south sides is quite reassuring, as one would not expect to have very different dust grains in different places of the disk (i.e., not surviving half an orbit). If indeed, there is an over-density of small dust grains along the north side as suggested in \citet{Olofsson2019}, then the slight bump at $90^{\circ}$ may not be real, and the true phase function may be more similar to the one of the south side.

\subsection{Other attempt at determining the phase function}

The approach presented in this study relies on some assumptions, for instance, how to estimate the surface brightness profiles of the model and of the observations. We originally attempted at least another approach that would have circumvented some of those issues, and we briefly discuss it here.

With the parameters of the disk fixed ($a$, $i$, etc), we sampled the phase function over a small number of angles ($10$), and tried to fit the actual values of the phase function, without any prior nor parametrization. The idea being that the code should find the shape of the phase function that minimizes the $\chi^2$. Unfortunately, the fitting never really converged. We postulate that the main issue with this approach is that we are trying to minimize second order effects. The $\chi^2$ is mostly dominated by the geometric shape of the ring, and small changes in the shape of the phase function yields very small changes in the final $\chi^2$ values. One possible work-around could be to work with relative $\chi^2$ values, but fine-tuning the evaluation of the goodness of fit may not be that trivial, and overall, we deemed this possible solution out of the scope of this paper.

Another alternative possibility, as mentioned in the introduction, would be to assume a parametric form for the phase function with a handful of free parameters (e.g., weighted sums of several Henyey-Greenstein functions, or a polynomial form). However, the challenge of such an approach would be to estimate when to increase the complexity of the form and when to stop. With our method, for a given set of disk parameters (e.g., $a$, $e$, $i$), the phase function that we retrieve is the best phase function that would minimize the goodness of fit (but it does not necessarily mean that it is a good solution).

\subsection{When does this matter?}

\begin{figure}
    \centering
    \includegraphics[width=\hsize]{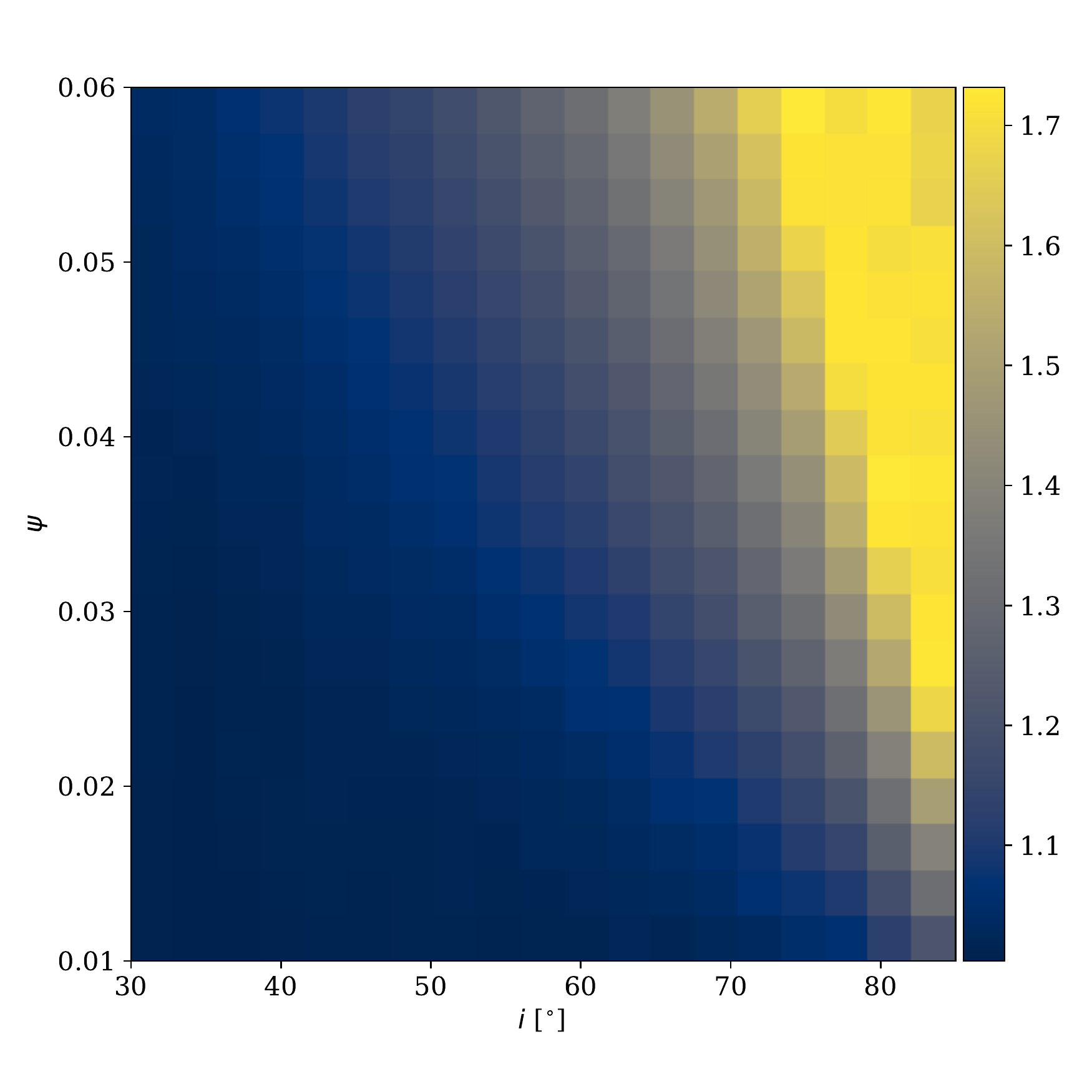}
    \caption{Map showing the surface brightness ratio between the major and minor axis of disk models with an isotropic phase function, as a function of the inclination and opening angle of the disk.}
    \label{fig:map}
\end{figure}

To quantify when the column density variations make a significant difference, we compute a grid of models using an isotropic phase function. We then compute the synthetic surface brightness as a function of the scattering angle similarly to the left panel of Figure\,\ref{fig:surface}. For each model we estimate the ratio between the maximum and minimum values of the profile. In the grid, we explore the two parameters that have the most important impact on the density increase; the inclination $i$ and opening angle $\psi$. The inclination ranges between  $30$ and $85^{\circ}$, and the opening angle ranges between $0.01$ and $0.06$ radians. The semi-major axis, position angle, $\alpha_\mathrm{in}$, and $\alpha_\mathrm{out}$ are set to the same values as before. To simplify the problem, we set $e=0$ (therefore $\omega$ no longer matter). Figure\,\ref{fig:map} shows the ratio of density enhancement between the major and minor axis of the disk for the grid. For disks that have an opening angle of $0.04$, the effect starts to become significant for inclinations larger than $\sim 60^{\circ}$.

\subsection{Is the disk around HR\,4796 vertically thin?}\label{sec:flat}

\begin{figure}
    \centering
    \includegraphics[width=\hsize]{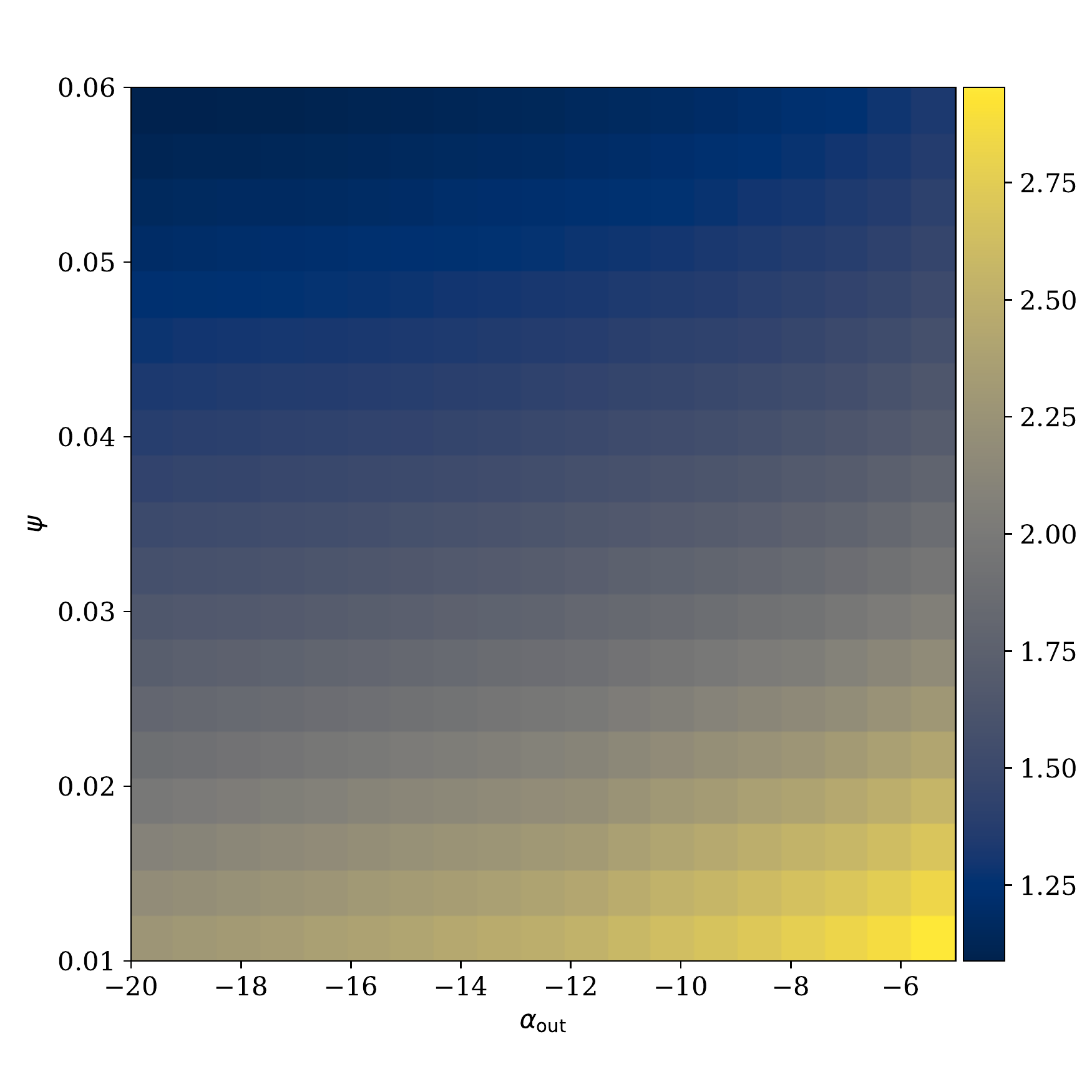}
    \caption{Map showing the ratio of the FWHM along the semi-major and semi-minor axis of disks models, as a function of the outer slope of dust density distribution and the opening angle of the disk.}
    \label{fig:height}
\end{figure}

\citet{Milli2019} injected the phase function they inferred into a disk model and obtained residuals that are comparable to the ones of Figure\,\ref{fig:data}, but assumed a vertically thin disk (vertical height of $1$\,au at a reference radius of $77.4$\,au). Following the description laid out in \citet{Augereau1999}, this would translate into an opening angle of $\sim 0.009$ for the code used in this work. Therefore, the true shape of the phase function depends on whether the disk is vertically flat or not.

\citet{Kennedy2018} modelled ALMA observations but could not firmly conclude on the vertical height of the disk. They mentioned that the disk could be vertically resolved, with a typical height of $\sim 10$\,au at a radius of $80$ \,au, but that a flat disk was also consistent with their observations. Given that a dynamical cold disk would be vertically thin (\citealp{Thebault2008}, but see also \citealp{Thebault2009}), and would explain its narrowness, the authors remained cautious and the actual vertical thickness of the disk remains a matter of debate. 

Nonetheless, \citet{Thebault2009} argued that debris disks should have a minimum aspect ratio $H/r$ of $0.04\pm0.02$ (where $H$ is the local half width at half maximum) and that disks are most likely stratified for different grain sizes; the smallest dust grains having a larger aspect ratio compared to larger grains. Therefore, debris disks may appear vertically thicker in scattered light observations than at millimetre wavelengths. They also argue that the disk vertical height cannot directly be related to its dynamical excitation. Converting our best-fit value for the opening angle to the aspect ratio as defined in \citet{Thebault2009}, we obtain a value of $0.041$, consistent with their results. 

For an inclined disk, the width measured along the major and minor axis should depend on the vertical thickness; if the disk is vertically flat the minor axis should appear narrower than the major axis due to the inclination, as illustrated in Figure\,\ref{fig:opang}. To better quantify this, we computed several models, with the same spatial resolution as the observations, varying the following two free parameters, the opening angle $\psi$ and the outer slope of the dust density distribution $\alpha_\mathrm{out}$ (which governs the width of the disk). For those models the semi-major axis and the inclination are the ones of the best fit model, and we used an isotropic phase function (in App.\,\ref{sec:app_pf} we repeat the same exercise for a different phase function to test the impact of this choice). Each image is convolved as explained in the previous Section. We measure the FWHM along the major and minor axis of the disk model and compute their ratio. Figure\,\ref{fig:height} shows how the ratio varies as a function of the two free parameters. While the slope of the dust density distribution has a small effect on the ratio of FWHM, the opening angle has the strongest impact. Overall, this suggests that the angular resolution of the observations is sufficient to constrain the height of the disk, and that this can in principle be done by measuring the width of the disk as a function of the azimuthal angle. 

However, this approach cannot be easily applied to our observations. The innermost regions are affected by strong noise, making the determination of the FWHM of the minor axis difficult. Furthermore, the background level is not the same along the minor and major axis of the disk, which may bias the peak values of both profiles, and hence the values of the FWHM. That being said, the test presented in Figure\,\ref{fig:height} strongly suggests that the width of the disk at different azimuthal angles informs us about its vertical structure. This supports our findings that the distribution of small dust grains in the disk around HR\,4796 is most likely vertically extended.

\subsection{Angular differential observations}\label{sec:adi}

We presented a new approach to estimate the phase function measured in polarimetric observations, but it would be extremely valuable to also be able to measure the phase function in total intensity from angular differential imaging (ADI) observations. The main challenge when modelling ADI observations is that self-subtraction effects cannot easily be dealt with \citep{Milli2012}. After median-collapsing the de-rotated cube (after performing principal component analysis, or any other algorithms), one cannot measure the surface brightness of the disk free of biases. Therefore we cannot properly estimate the surface brightness of the disk to correct the phase function for column density effects. 

\begin{figure}
\centering
\includegraphics[width=\hsize]{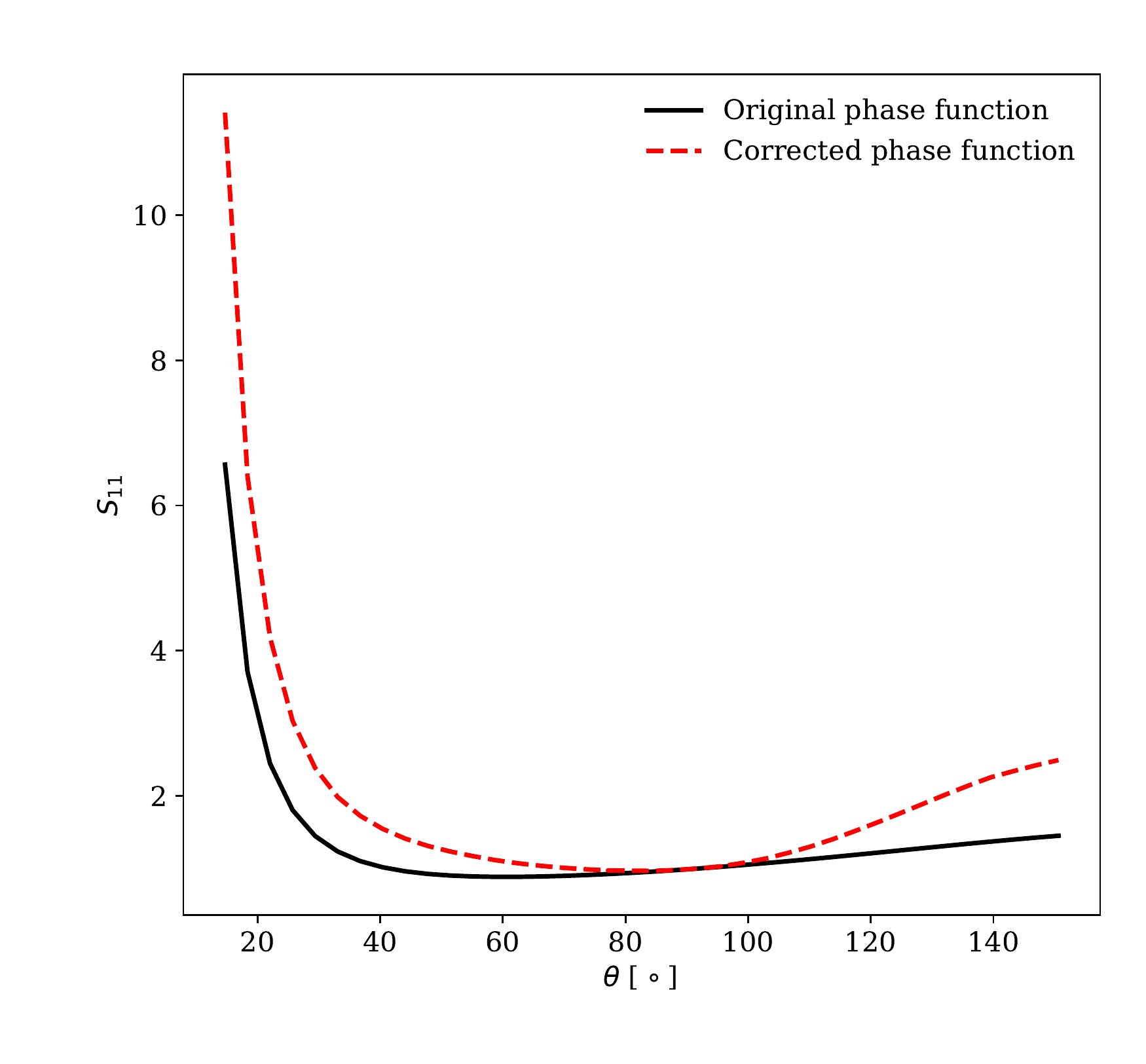}
    \caption{Fit to the total intensity phase function derived in \citet{Milli2017} using two Henyey-Greenstein functions (black) ``corrected'' for projection effects (dashed red). Both functions are normalized to unity at $90^{\circ}$.}
\label{fig:s11}
\end{figure}

One way to avoid biases induced by self-subtraction is to perform ``forward modelling'' by subtracting a disk model in the cube (and we then need another free parameter to scale up or down the model before subtracting it). Having the scattered light phase function as a free parameter (as mentioned in a previous sub section) runs into the same shortcomings. One possible, but time costly, approach would be to run a disk model with an isotropic total intensity phase function, find the best scaling factor for the given set of disk parameters, measure the residuals in the final image and modify the shape of the phase function accordingly before re-evaluating the scaling factor.

Nonetheless, we here attempt to roughly quantify the changes to the phase function measured on total intensity observations of HR\,4796. \citet{Milli2017} presented SPHERE/IRDIS observations of the disk, and measured the phase function from the surface brightness distribution, correcting for self-subtraction effects based on a disk model. This model has parameters that are compatible with our best fit solution, with a slightly different value for $e$ ($0.06$) and therefore for $\omega$ as well ($-105.7^{\circ}$ in our reference frame) given the degeneracy between the two parameters. They then fitted two weighted Henyey-Greenstein functions $f_\mathrm{HG}$ to the measured phase function, as $w \times f_\mathrm{HG}(g_0) + (1-w) \times f_\mathrm{HG}(g_1)$, where $f_\mathrm{HG}$ takes the following form
\begin{equation}
    f_\mathrm{HG}(g) = \frac{1}{4 \pi} \frac{1-g^2}{(1 + g^2 - 2g\mathrm{cos}\theta)^{3/2}}.
\end{equation}
\citet{Milli2017} found that $g_0 = 0.99$, $g_1 = -0.14$, and $w = 0.83$. If the disk is not flat, the phase function they inferred does not take into account column density variations. We therefore used their analytical form, and applied an additional correction factor based on the dust density variation as a function of the scattering angle for a non-flat disk. The revised phase function being the original phase function divided by the surface brightness of the best fit model with an isotropic phase function. Both phase functions are shown in Figure\,\ref{fig:s11}, normalized to unity at $90^{\circ}$. The most notable difference is that backward scattering becomes much more significant when accounting for the column density variations due to the inclination. One should keep in mind however that this result remains model dependent for both the self-subtraction and the density variation corrections (and the discussion of Section\,\ref{sec:flat}).

Interestingly, as a side note, \citet{Ren2019} measured the surface brightness of the disk and halo around HD\,191089 and measured strong backward scattering for the halo but not for the main ring. If the halo is vertically very thin, extending mostly from the densest regions, i.e., the midplane, then one should not expect significant column density changes due to the inclination of the system ($\sim 59^{\circ}$) and therefore the measurement of the phase function from the halo would not suffer from the same biases described in this paper.

\subsection{Dust properties in the disk around HR\,4796}

\begin{figure*}
\centering
\includegraphics[width=\hsize]{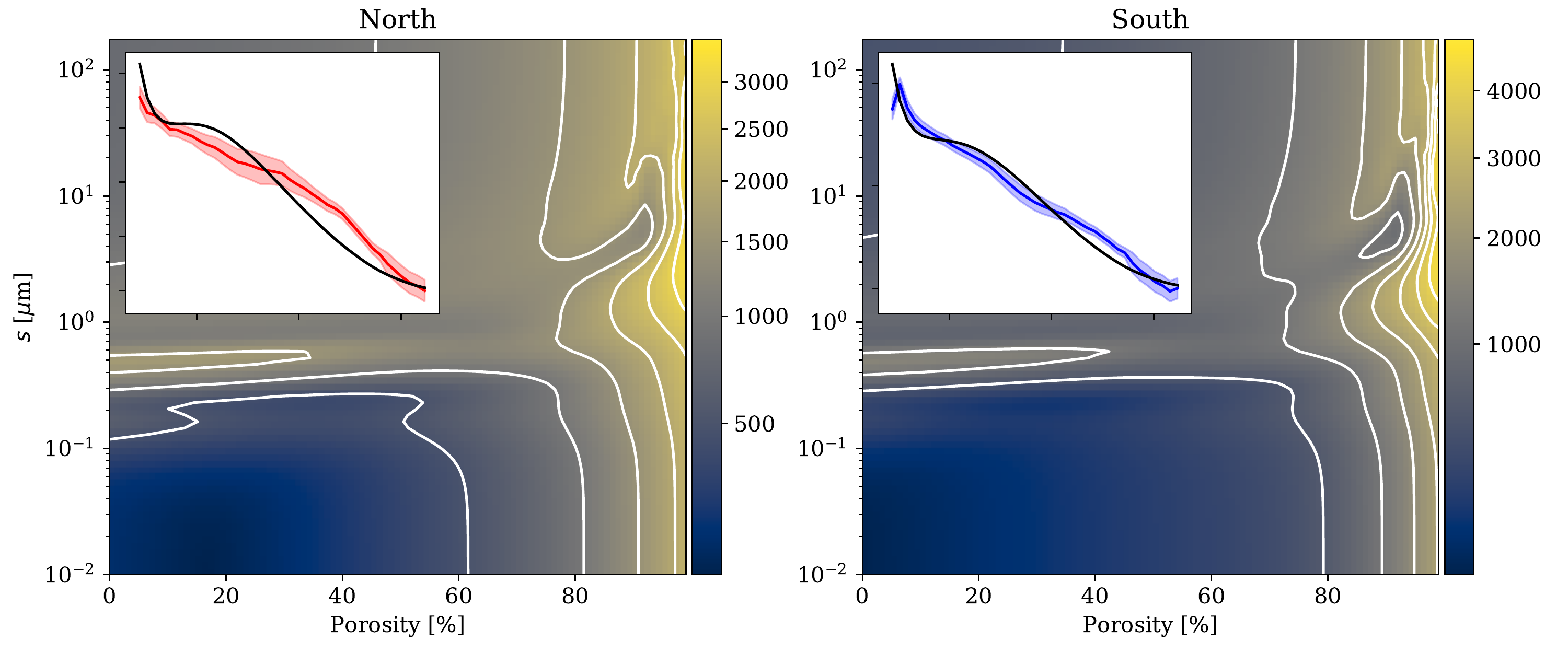}
\caption{Maps of the $\chi^2$ when trying to reproduce the shape of the phase functions using DHS for the north and south sides (left and right, respectively). The two free parameters are the porosity and the minimum grain sizes. The insets show the observations and the best-fitting model}
\label{fig:dustprop}
\end{figure*}

With the revised phase functions, we can attempt to revise the dust properties inferred in \citet{Milli2019}. We computed a grid of polarized phase functions, using the \texttt{OpacityTool} (\citealp{Woitke2016}, \citealp{Toon1981}). The code can compute absorption and scattering properties, as well as six elements of the scattering matrix, including $S_{11}$ and  $S_{12}$, assuming a distribution of hollow spheres (DHS, \citealp{Min2005}). We assume that the grain size distribution is a differential power-law of the form d$n(s) \propto s^{-3.5}$d$s$, where $s$ is the grain size, and we set $s_\mathrm{max} =  1$ \,mm. We compute the polarized phase function, integrated over the size distribution, varying $s_\mathrm{min}$ between $0.01$\, $\mu$m and $110\,\mu$m, and the porosity between $0$ and $99$\%. The optical constant are taken from \citet[][amorphous silicates]{Dorschner1995} with a $85$\% mass fraction and \citet[][amorphous carbon]{Zubko1996} with a $15$\% mass fraction. The maximum filling factor is set to $0.8$. To estimate whether we can reproduce the inferred phase functions, we computed a grid of models, and the $\chi^2$ maps for both the north and south sides are shown in Figure\,\ref{fig:dustprop} (left and right, respectively). For both panels, the insets show the inferred phase functions and their best-fit model. For both sides, we find that the best model is obtained for $s_\mathrm{min} = 0.01$ \,$\mu$m. For the north side, we find that the porosity should be $16$\%, and $0$\% for the south side, with significant uncertainties as illustrated in Fig.\,\ref{fig:dustprop}. In Appendix\,\ref{sec:poro_size} we show in more detail the effects of porosity and minimum grain size on the shape of the phase function. 

While the best-fit model reproduces rather well the phase function for the south side, it fails to capture the shape of the north side. One has to keep in mind that, as discussed before, the northern phase function may be biased by a possible over-density of small dust grains, but overall, we find that the phase functions suggest the presence of very small dust grains, with rather low porosity values.

With the exercise described above, the slope of the grain size distribution is set to $-3.5$, to minimize the number of free parameters, but which may be a strong hypothesis. The grain size distribution can show some wavy structures (e.g., \citealp{Thebault2007}) or there can be an over-abundance of small dust grains in bright debris disks (as is the case for HR\,4796, \citealp{Thebault2019}). Therefore, we repeated the same exercise as before, but integrating the size distribution between $s$ and $s+\delta_s$, keeping the slope as $-3.5$. The motivation being to identify the characteristic size that can best explain the phase functions.  We use \texttt{PyMultiNest} to find the best fit model and for the north side, we obtain $s_\mathrm{min} = 0.03\pm0.02$\,$\mu$m, $\delta_s = 0.29\pm0.15$\,$\mu$m, and a porosity of $16\pm1$\%. For the south side, we find $s_\mathrm{min} = 0.02\pm0.01$ \,$\mu$m, $\delta_s = 0.36\pm0.29$\,$\mu$m, and a porosity compatible with $0$\%. The shape of the best fit model is quite similar to the previous best fit models. The results of this approach are quite similar to the previous ones, the phase functions are best reproduced by very small dust grains. 

Using the \texttt{OpacityTool}, we can also compute the asymmetry parameter $g_\mathrm{sca}$ and compute the unitless $\beta$ ratio between the stellar radiation pressure and the gravitational force for different grain sizes, as
\begin{equation}
    \beta(s) = \frac{3L_\star}{16\pi G c^2 M_\star}\frac{Q_\mathrm{pr}(s)}{\rho s},
\end{equation}
where $L_\star$ and $M_\star$ are the stellar luminosity and mass ($25.75$\,$L_\odot$ and $1.31$\,$M_\odot$, respectively, \citealp{Olofsson2019}), $G$ the gravitational constant, $c$ the speed of light, $\rho$ the dust density, and $Q_\mathrm{pr}$ the radiation pressure efficiency (equal to $Q_\mathrm{ext}(\lambda, s) - g_\mathrm{sca}(s) \times Q_\mathrm{sca}(\lambda, s)$) averaged over the stellar spectrum. For porosity values of $0$ and $16\%$ we find that the blow-out sizes (for which $\beta \leq 0.5$, assuming the parent bodies are on circular orbit) are $\sim 13.5$ and $17$\,$\mu$m (for larger porosity values, the blow-out size increases even more). All the grains that are smaller no longer are bound to the star and would be removed from the system rapidly. We therefore reach the same conclusions as the ones presented in \citet{Milli2017,Milli2019}, that the Mie or DHS theory cannot adequately explain the full picture. Indeed, \citet{Augereau1999} found that to reproduce the spectral energy distribution of the disk, the minimum grain size should be close to $10$\,$\mu$m, which is rather compatible with the aforementioned blow-out size but would fail to reproduce the measured phase function. Relatively large aggregates composed of sub-$\mu$m sized monomers may be a viable alternative to explain the observations. As explained in \citet{Min2016}, the polarization properties of aggregates are intimately related to the size of the individual monomers and not to the overall size of the aggregate itself.

\section{Conclusions}

In this paper, we presented an alternative approach to estimate the phase function from polarized observations of debris disks with an emphasis on disks that have a non negligible vertical scale height. While our method remains model-dependent it does not require a parametrized form for the phase function (e.g., Henyey-Greenstein). The total flux depends both on the local density and the true phase function, but when the disks are highly inclined, and not infinitely flat, there are variations in column density along the major axis, due to projection effects, and those variations have to be taken into account.

We presented an approach to account for those column density variation effects, and find that the inferred phase function is quite different from previous estimates. We tested our model to SPHERE/ZIMPOL observations of HR\,4796 and derived phase functions for both the north and south sides, both being relatively similar. We reach similar conclusions as the ones outlined in \citet{Milli2019}, i.e., we cannot fully reconcile all key aspects with a single scattering theory (e.g., phase function and blow-out size). Our modelling results suggest that the disk is not vertically flat, with an opening angle of $\psi \sim 0.035$. The vertical scale height can successfully be constrained by the model based on how the width of the disk varies as a function of the azimuthal angle. 

We also note that our modelling approach remains blind to any local increase of the dust density, and that it cannot readily be applied to ADI observations. We remark however that similar biases are probably occurring when deriving the total intensity phase function, which may lead to an under-estimation of backward scattering.

\begin{acknowledgements}
We are grateful to the referee, Kees Dullemond, for providing helpful comments, especially with respect to the determination of the vertical scale height, as well as several pointers to help clarify the paper.
This research has made use of the SIMBAD database (operated at CDS, Strasbourg, France). This research made use of Astropy, a community-developed core Python package for Astronomy (\citealp{Astropy}).
J.~O. and A.~B. acknowledge support from the ICM (Iniciativa Cient\'ifica Milenio) via the Nucleo Milenio de Formación planetaria grant. J.~O. acknowledges support from the Universidad de Valpara\'iso and from Fondecyt (grant 1180395). A.~B. acknowledges support from Fondecyt (grant 1190748). 
\end{acknowledgements}

\bibliographystyle{aa}

\appendix

\section{Miscellaneous}

\subsection{Corner plot}

\begin{figure*}
\centering
\includegraphics[width=\hsize]{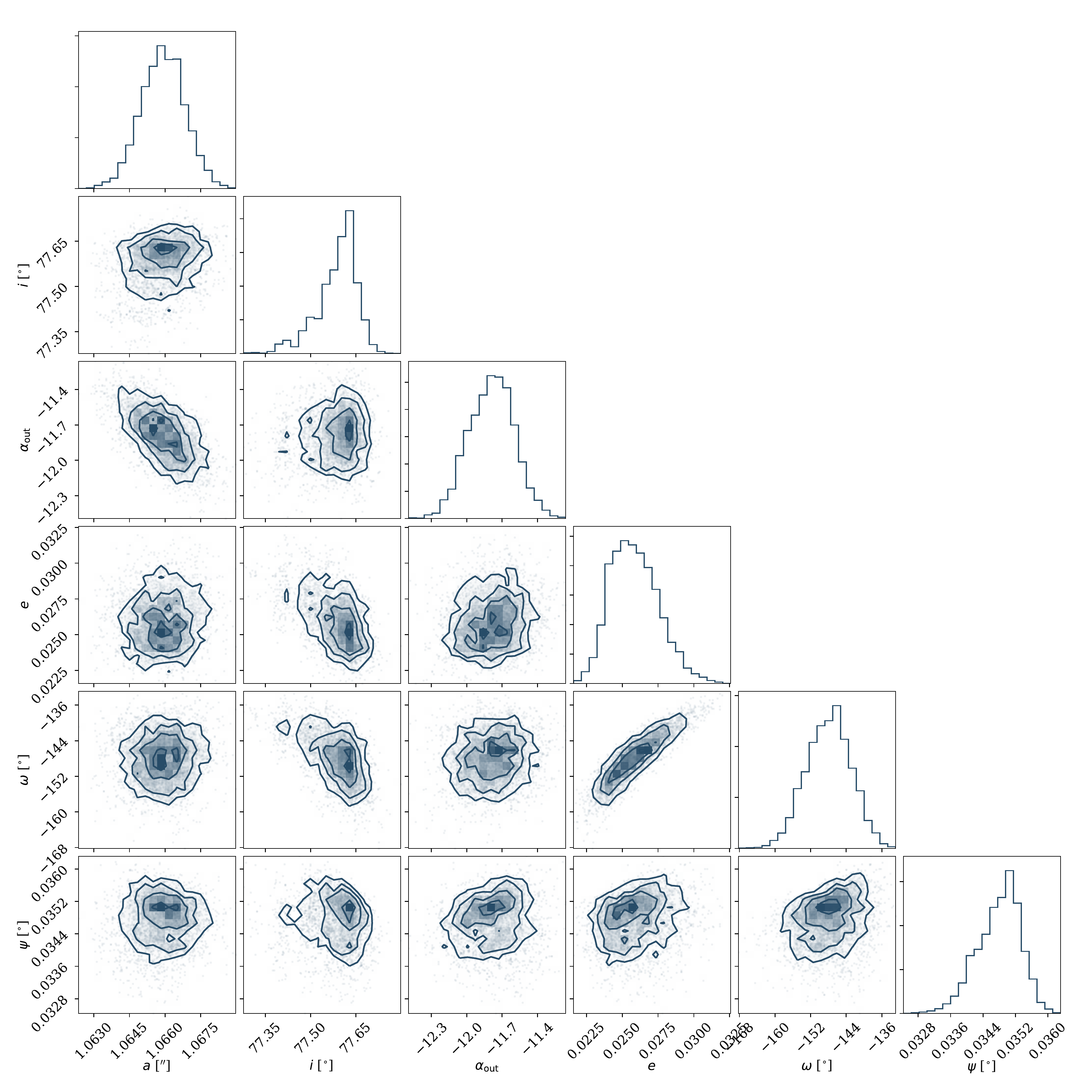}
\caption{Corner plot of the posterior density distributions for the modelling.}
\label{fig:corner}
\end{figure*}

Figure\,\ref{fig:corner} shows the corner plot for the modelling, with the density plots and the projected probability distributions for each parameter. 

\subsection{The impact of the phase function on the apparent width of the disk}\label{sec:app_pf}

\begin{figure}
    \centering
    \includegraphics[width=\hsize]{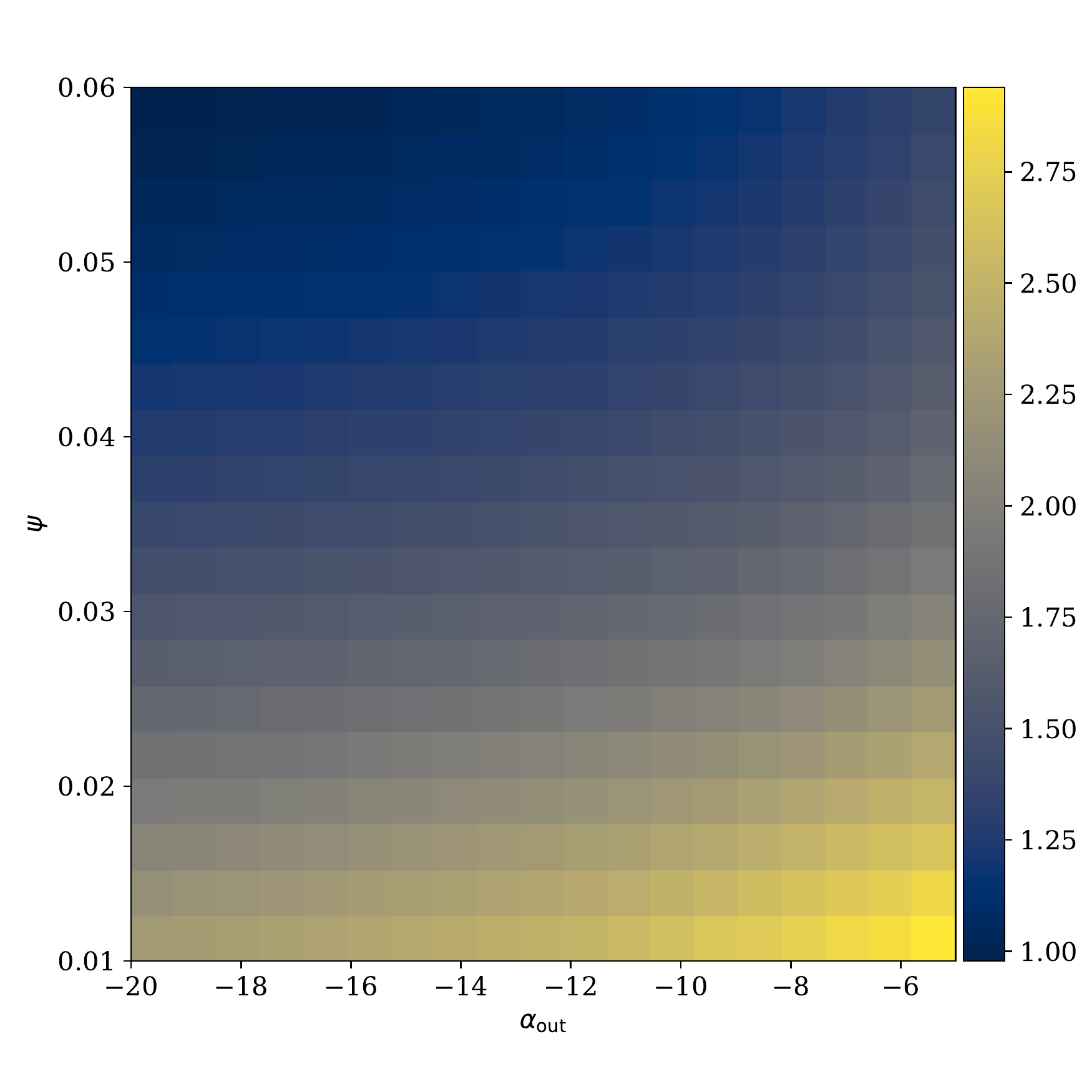}
    \caption{Same as Figure\,\ref{fig:height} using a different phase function for the models.}%
    \label{fig:height09}
\end{figure}

In Section\,\ref{sec:flat} we computed the ratio of FWHM along the major and minor axis of disk models to assess whether the width of the disk at different azimuthal angles can inform us about its vertical scale height. In Figure\,\ref{fig:height}, we used an isotropic phase function, but the phase function can change the intensity along the major and minor axis, and as a consequence the measure of the FWHM. Therefore, we here repeat the same exercise but with a different phase function. We computed models in total intensity (and not polarized light) using the Henyey-Greenstein approximation and with a $g$ value of $0.9$. This choice is motivated by the fact that the phase function strongly peaks at small scattering angles, significantly enhancing the intensity along the minor axis. Therefore, this phase function and the isotropic one are very complementary ones, allowing us to further assess the robustness of our findings. The fact that we are computing total intensity images and not polarimetric images is not relevant for the interpretation of the results. We proceeded in the same way as described in Section\,\ref{sec:flat}, computing the models, convolving them and measuring the two FWHM. The results are presented in Figure\,\ref{fig:height09} and the results are compatible with Figure\,\ref{fig:height} with comparable values for the ratio. We therefore conclude that while the choice of the phase function as a small impact on the appearance of the disk, it is of second order compared to the effect of the density distribution. 

\subsection{Porosity and grain size}\label{sec:poro_size}

\begin{figure*}
\centering
\includegraphics[width=0.45 \hsize]{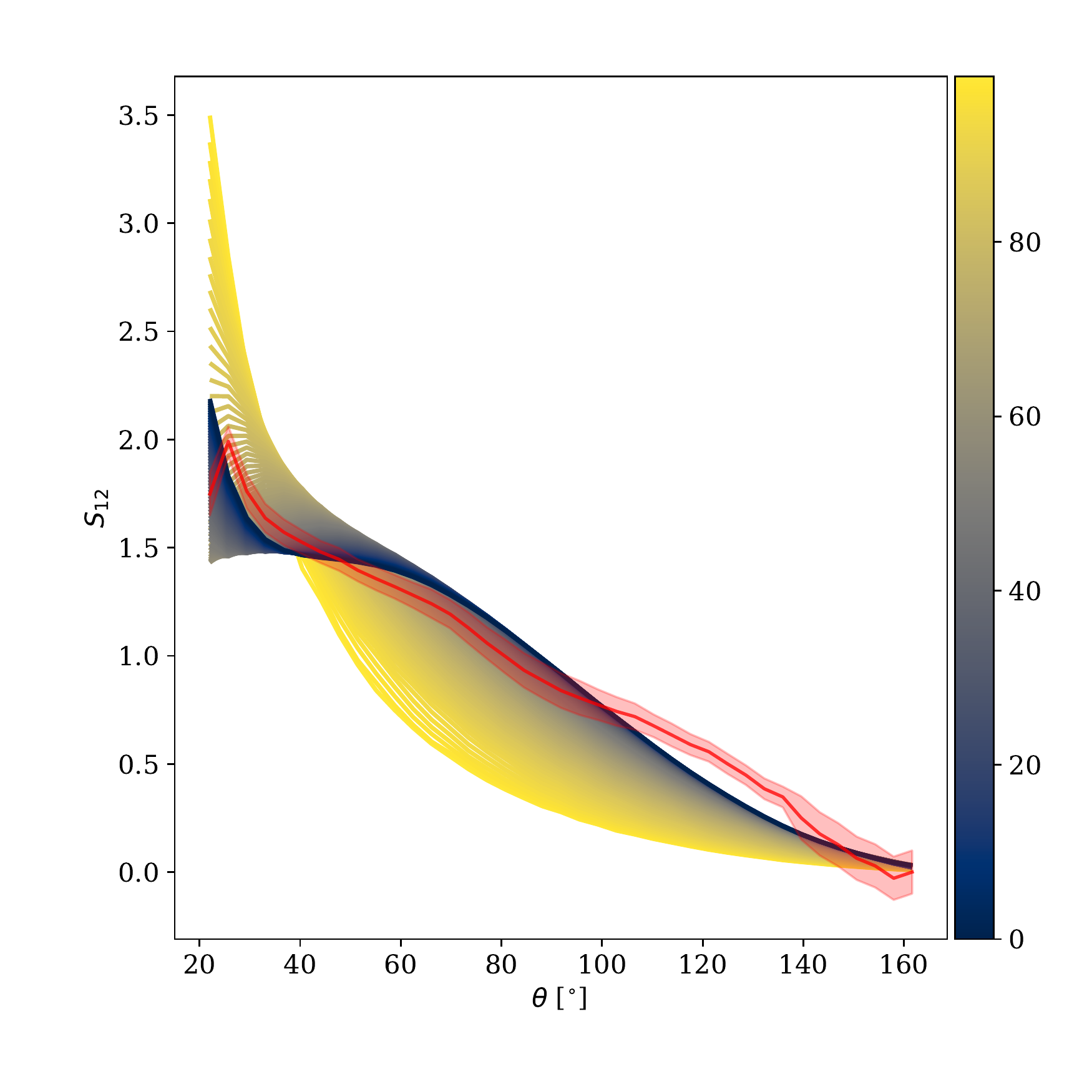}
\includegraphics[width=0.45 \hsize]{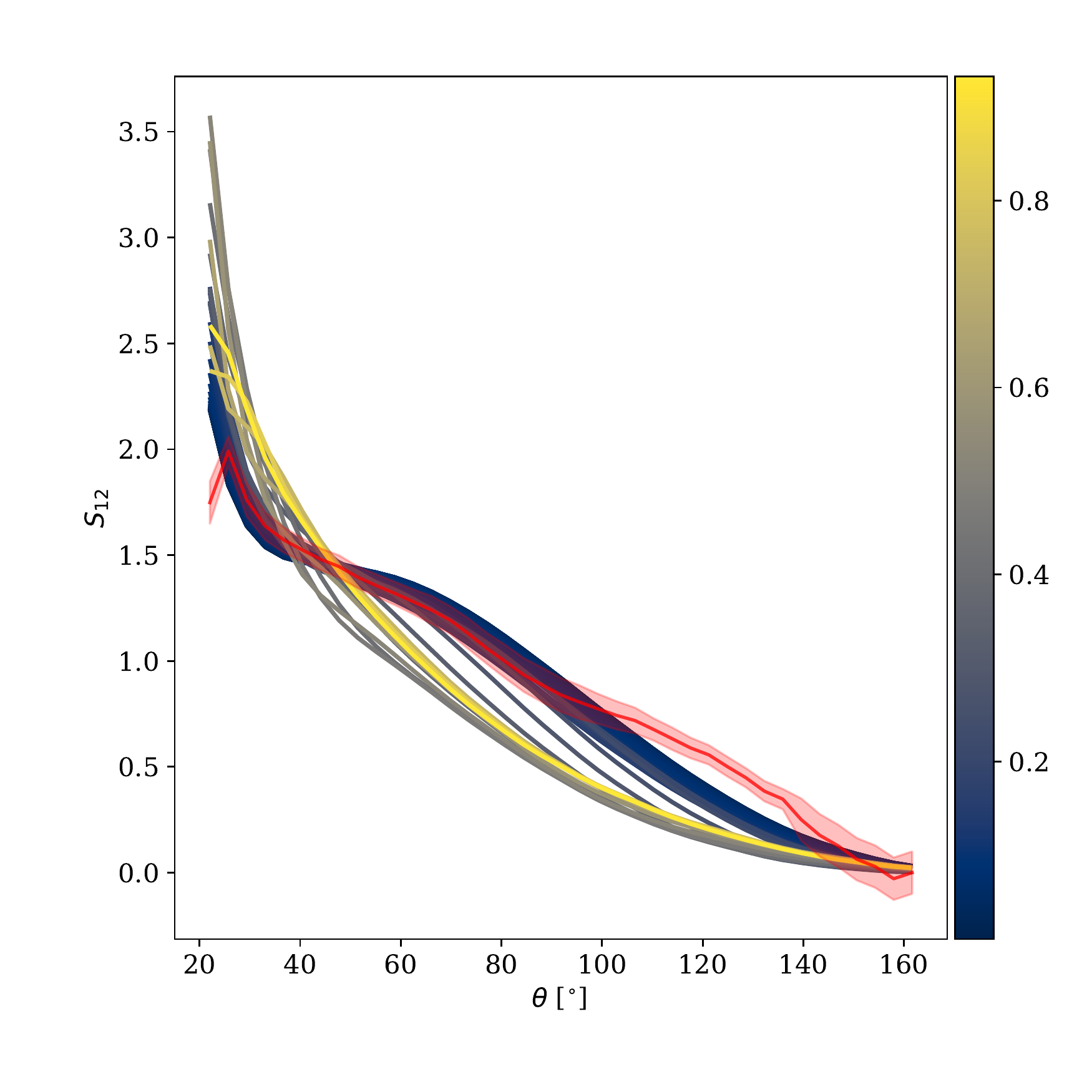}
    \caption{Different phases functions compared to the phase function estimated for the south side of the disk, for different values of the porosity (left) and minimum grain size (right). The color bars are in units of percent and $\mu$m (left and right, respectively).}
\label{fig:poro_size}
\end{figure*}

To complement Figure\,\ref{fig:dustprop}, Figure\,\ref{fig:poro_size} shows the effect of the porosity (left panel) and minimum grain size (right panel) on the phase function, compared to the phase function of the south side of the disk, to highlight how the shape varies as a function of those two parameters.

\end{document}